\documentclass[aps,prl,twocolumn,showpacs,groupedaddress]{revtex4}

\usepackage{graphics,bm}
\usepackage{graphicx}

\newcommand{\beq}{\begin{equation}}
\newcommand{\eeq}{\end{equation}}
\newcommand{\bqa}{\begin{eqnarray}}
\newcommand{\eqa}{\end{eqnarray}}
\def\gsim{\mathrel {\vcenter {\baselineskip 0pt \kern 0pt
\hbox{$>$} \kern 0pt \hbox{$\sim$} }}}

\protect

\begin{document}
\title{Exact Solitonic Solutions of the Gross-Pitaevskii Equation with a Linear Potential}
\author{Usama Al Khawaja}
\affiliation{ \it Physics Department, United Arab Emirates
University, P.O. Box 17551, Al-Ain, United Arab Emirates.}

\date{\today}

\begin{abstract}

We derive classes of exact solitonic solutions of the time-dependent
Gross-Pitaevskii equation with repulsive and attractive interatomic interactions.
The solutions correspond to a string of bright solitons with phase
difference between adjacent solitons equal to $\pi$. While the
relative phase, width, and distance between adjacent solitons turn
out to be a constant of the motion, the center of mass of the string
moves with a constant
acceleration arising from the inhomogeneouty of the background.

\end{abstract}


\maketitle

{\it Introduction}---The experimental realization of dark solitons
\cite{burger,denschlag,anderson}, bright solitons
\cite{randy,schreck}, and recently gap solitons \cite{eirman} in
Bose-Einstein condensates has stimulated intense interest in their
properties particularly their formation and propagation
\cite{linear1,linear2,cast,sala,busch,abdu}. Due to the nonlinearity
arising from the interatomic interactions and due to the presence of
a confining potential in the Gross-Pitaeviskii equation that
describes the evolution of the solitons, these studies were
performed either by solving the corresponding Gross-Pitaevskii
equation numerically or by using perturbative methods. Interestingly
enough, some exact solitonic solutions were recently found for this
equation in one dimension, but with time- and space-dependent
interatomic interactions and trapping potential strengths
\cite{wu,jun,liang,lu,raj}. Obtaining such exact solutions allows
for testing the validity of the Gross-Pitaevskii equation at high
densities, obtain the long-time evolution of the soliton where
numerical techniques may fail, and helps to understand soliton
formation and propagation.

The so-called Darboux transformation method \cite{salle} was used to
obtain such exact solutions. We have shown in a previous work
\cite{usama_darboux} that such exact solutions may exist only for
specific functional forms for the interatomic interaction and trapping
potential strengths. For example, in the work of Liang {\it et al.}
\cite{liang}, the exact solution is found only when the trapping
potential is quadratic, expulsive, and the interatomic
interaction strength is growing exponentially with time with a rate
that equals the trapping potential strength itself. Such a
restriction, makes the exact solution less interesting from an
experimental point of view. In an attempt to soften this
restriction, we found that exact solutions may also be found for
constant, linear, or quadratic potentials, and with interatomic
interaction strengths that can be constant, growing, or decaying in
time \cite{usama_darboux}.

Here, we exploit our previous result to obtain exact solitonic
solutions of the time-dependent Gross-Pitaevskii equation with
linear trapping potential and constant interatomic interaction
strength that can be positive or negative. The Gross-Pitaevskii
equation describes, in this case, the surface of the condensate.

{\it The Gross-Pitaevskii equation.}--- Near the surface of a
Bose-Einstein condensate, the quadratic trapping potential can be
approximated by a linear potential and the surface of the condensate
can be regarded as an infinite plane. The Gross-Pitaevskii equation,
in this case, takes the form \cite{emil,usama_peth}
\begin{equation}
i\hbar{\partial\psi(x,t)\over\partial t}= \left[
-{\hbar^2\over2m}{\partial^2\over\partial x^2} +F x+{4\pi a
\hbar^2\over m}|\psi(x,t)|^2\right]\psi(x,t) \label{gp1},
\end{equation}
where $\psi(x,t)$ is the condensate wavefunction, and $x$ is the
coordinate normal to the surface of the condensate such that the
bulk of the condensate exists in the region $x<0$. The force
constant $F$ arises from linearizing the harmonic potential near the
surface, namely $F=m\omega^2 R$, where $\omega$ is the
characteristic frequency of a spherically symmetric harmonic
trapping potential, $R$ is the radius of the condensate, and $m$ is
the mass of an atom. The interatomic interaction strength is
proportional to the scattering length $a$ which can be positive or
negative.

The characteristic length $\delta$ in the surface region is defined
by equating the kinetic energy $\hbar^2/2m\delta^2$ to the potential
energy $F\delta$, namely $\delta=(\hbar^2/2m F)^{1/3}$. Scaling
length to $\delta$, time to $\tau=2m\delta^2/\hbar$, and the
wavefunction to $\sqrt{\rho_0}=1/\sqrt{8\pi|a|\delta^2}$, the
previous equation takes the dimensionless form
\begin{equation}
i{\partial\psi(x,t)\over\partial t}= \left[
-{\partial^2\over\partial x^2} +x-p^2|\psi(x,t)|^2\right]\psi(x,t)
\label{gp2},
\end{equation}
where we have retained the nonscaled symbols for convenience. The
parameter $p^2=-{\rm sgn}(a)$ allows for treating the repulsive case
($p^2=-1$) and attractive case ($p^2=1$) simultaneously.

For the case of repulsive interactions, the Thomas-Fermi
approximation can be used to estimate $\delta$ and $\rho_0$ in terms
of the size of the condensate, $R$, and the central density
$\rho_{\rm TF}$ \cite{usama_thesis}. It turns out that
$\delta/R\approx\gamma^{-4/15}$ and $\rho_0/\rho_{\rm
TF}\approx\gamma^{-4/15}$, where  $\gamma=Na/a_0$ is the
dimensionless interaction strength, $a_0=\sqrt{\hbar/m\omega}$ is
the characteristic length of the harmonic oscillator potential, and
$N$ is the number of atoms. For a typical $^{87}$Rb condensate with
$10^4$ atoms the two ratios are roughly equal to 1/4. The unit of
time is, in this case, given by $\tau\approx2\gamma^{-1/15}/\omega$.
Approximating the quadratic potential of the Bose-Einstein condenste
by a linear one in the surface region is accurate only within a
region of width $\delta$ around the {\it Thomas-Fermi} surface. It
turns out, however, that some of the solitonic solutions we obtain
here have width larger than $\delta$. Furthermore, the dynamics of
these solitons is such that they drift from the surface region
towards the bulk region of the condensate where the potential is not
linear anymore. Therefore, such solutions can be considered only as
the initial states of the time-dependent solitonic excitations of
the condensate.

{\it The Darboux Transformation and the New Solutions.}--- The first
step in the Darboux transformation method is to find a linear system
of equations for an auxiliary field ${\bf \Psi}(x,t)$ such that
Eq.~(\ref{gp2}) is its consistency condition \cite{salle}. Using the
method described in Ref.~\cite{usama_darboux}, we find that the
following linear system corresponds to Eq.~(\ref{gp2})
\begin{equation}
{\bf \Psi}_x={\bf J}\Psi{\bf\Lambda}+{\bf U}\Psi \label{psi_x},
\end{equation}
\begin{equation}
i{\bf \Psi}_t= {\bf W}\Psi+2(\zeta{\bf J}+{\bf U}){\bf
\Psi}{\bf\Lambda}+2{\bf J}\Psi{\bf\Lambda}^2 \label{psi_t},
\end{equation}
where,
\begin{math}
{\bf \Psi}(x,t)=\left(\begin{array}{cc}
\psi_1(x,t)&\psi_2(x,t)\\
\phi_1(x,t)&\phi_2(x,t)
\end{array}\right)
\end{math},
\hspace{0.25cm}
\begin{math}
{\bf J}=\left(\begin{array}{cc}
1&0\\
0&-1
\end{array}\right)
\end{math},
\hspace{0.25cm}
\begin{math}
{\bf \Lambda}=\left(\begin{array}{cc}
\lambda_1&0\\
0&\lambda_2
\end{array}\right)
\end{math},
\hspace{0.25cm}
\begin{math}
{\bf U}=\left(\begin{array}{cc}
\zeta&p\,q(x,t)/\sqrt{2}\\
-p\,r(x,t)/\sqrt{2}&-\zeta
\end{array}\right)
\end{math},
\\
\begin{math}
{\bf W}=(\zeta^2-x/2){\bf J}+2\zeta{\bf U}-{\bf J}({\bf U}^2-{\bf
U}_x)
\end{math}, $\zeta(t)=it/2$, and $\lambda_1$ and $\lambda_2$ are arbitrary constants.
The subscripts $x$ and $t$ denote partial derivatives with respect
to $x$ and $t$, respectively. Equation~(\ref{gp2}) is obtained from
the consistency condition $\Psi_{xt}=\Psi_{tx}$ and by substituting
$q(x,t)=r^*(x,t)=\psi(x,t)$.

This linear system of 8 equations, Eqs.~(\ref{psi_x}) and
(\ref{psi_t}), reduces to an equivalent system of 4 equations with
nontrivial solutions by making the following substitutions:
$\lambda_1=-\lambda_2^*$, $\phi_1=\psi_2^*$, $\psi_1=-p^2\phi_2^*$.
The reduced system can be solved once the so-called {\it seed}
solution $r(x,t)=q^*(x,t)$ is specified. We have also shown in
Ref.~\cite{usama_darboux} that the wavefunction
$\psi_0(x,t)=A\exp{(i\phi_0)}$ with $\phi_0=t(p^2A^2-(t^2/3+x))$,
where A is a real constant, is an exact solution of Eq.~(\ref{gp2}),
and thus can be taken as the seed solution.

The Darboux transformation can now be applied to the linear system
to generate a new solution of Eq.~(\ref{gp2}) as follows
\cite{salle}
\begin{equation}
\psi(x,t)=\psi_0(x,t)-{\sqrt{8}\over
p}(\lambda_1+\lambda_1^*)\phi_2\psi_2^*/(p^2|\phi|^2+|\psi|^2)
\label{psinew}.
\end{equation}
Substituting for $\psi_0(x,t)$, $\psi_2(x,t)$, and $\phi_2(x,t)$, we
obtain the following new exact solutions to Eq.~(\ref{gp2}): For the
repulsive interactions case ($p=\pm i$), the solution is
\begin{widetext}
\begin{equation}
\psi(x,t)=e^{i\phi_0}\left[ A\pm i{\sqrt{8}\lambda_{1r}}\,{
2u_r^+\cosh{\theta} -2iu_i^+\sinh{\theta} +(|u^+|^2+1)\cos{\beta}
+i(|u^+|^2-1)\sin{\beta} \over(|u^+|^2-1)\sinh{\theta
}+2u_i^+\sin{\beta} } \right] \label{psi_rep},
\end{equation}
and for the attractive interactions case ($p=1$), the solution is
\begin{equation}
\psi(x,t)=e^{i\phi_0}\left[ A-{\sqrt{8}\lambda_{1r}}\times{
2u_r^+\cosh{\theta} -2iu_i^+\sinh{\theta} +(|u^+|^2+1)\cos{\beta}
+i(|u^+|^2-1)\sin{\beta} \over(|u^+|^2+1)\cosh{\theta
}+2u_r^+\cos{\beta} } \right] \label{gp_attr},
\end{equation}
\end{widetext}
where \begin{math}
\theta=\sqrt{2}\left[\Delta_r(t^2+x)+2(\Delta_r\lambda_{1i}-\Delta_i\lambda_{1r})t\right]-\delta_r
\end{math},
\begin{math}
\beta=-\sqrt{2}\left[\Delta_i(t^2+x)+2(\Delta_i\lambda_{1i}+\Delta_r\lambda_{1r})t\right]+\delta_i
\end{math},
$u^\pm=\sqrt{8}\,p\,A/b^\pm$, $b^\pm=4\lambda_1^*\pm\Delta$,
$\Delta=\sqrt{2{\lambda_1^*}^2-p^2A^2}$, and $\delta$ is an
arbitrary constant. Here, the subscripts $r$ and $i$ denote real and
imaginary parts, respectively.

\begin{figure}
\begin{center}
\includegraphics[width=9cm]{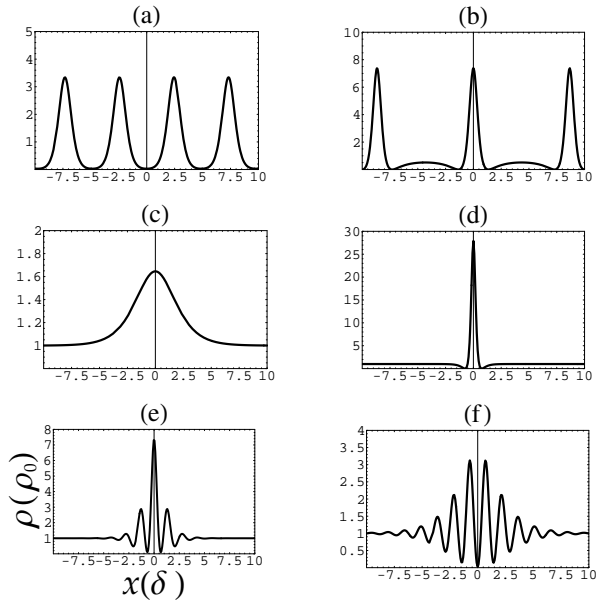}
\end{center}
\caption{Density $\rho(x)=|\psi(x)|^2$ at time $t=0$ for the case of
attractive interactions. The arbitrary constants chosen to generate
these plots are: $\delta_r=\delta_i=0$ and $A=1$ for all plots. In
(a) $\lambda_{1i}=0$, $\lambda_{1r}=0.29$, in (b): $\lambda_{1i}=0$,
 $\lambda_{1r}=-0.6$, in (c): $\lambda_{1i}=0$,
$\lambda_{1r}=0.8$, in (d): $\lambda_{1i}=0$, $\lambda_{1r}=-1.5$,
in (e): $\lambda_{1i}=2$, $\lambda_{1r}=-0.6$, and in (f):
$\lambda_{1i}=2$, $\lambda_{1r}=0.29$. Typical values of the length
and density units, $\delta$ and $\rho_0$, are given in the text.}
\label{fig1}
\end{figure}

{\it Properties of the Solutions.}--- Here, we describe the main
properties and features of the exact solitonic solutions found
above. The five arbitrary constants $\delta_r$, $\delta_i$,
$\lambda_{1i}$, $\lambda_{1r}$ and $A$, have the following effects
on the solutions: The constants $\delta_r$ and $\delta_i$ have the
trivial effect of shifting the solutions in the $x$- and
$t$-coordinates. Therefore, we set from now on
$\delta_r=\delta_i=0$. The other three constants are combined under
a square root in the above expression for $\Delta$. Thus, depending
on the values of these constants, $\Delta$ can be real, imaginary,
or complex. If $\Delta$ is real, $\beta$ will have no $x$-dependence
and the solutions will be nonoscillatory, i.e., single-soliton
solution. If $\Delta$ is imaginary, $\theta$ will have no
$x$-dependence, and the solution is oscillatory. If $\Delta$ is
complex, the solution will be a combination of both previous cases,
namely oscillatory but with a localized envelope which has a density
profile that is similar to that of {\it gap} solitons (See Fig.1 of
Ref.~\cite{beata}). These cases are shown in In Table 1, and
specific cases are visualized in Fig.~\ref{fig1}, where we plot the
density $\rho(x,t)=|\psi(x,t)|^2$ at $t=0$ for the attractive
interactions case.

\begin{table}
\begin{tabular}{|l|l|l|}
\hline
&$\lambda_{1r}>0$&$\lambda_{1r}<0$\\
\hline \hline$\lambda_{1i}=0$& multi-solitonic&multi-solitonic\\
& with broad edges &with sharp edges\\
\hline\hline
$\lambda_{1i}<0$&single-solitonic&single-solitonic\\
&with broad edges&with sharp edges \\
\hline\hline
$\lambda_{1i}>0$&multi-solitonic&multi-solitonic\\
&with envelope&with envelope\\
\hline
\end{tabular}
\caption{Classification of the solitonic solutions. Fig.1a shows an
example of the multi-solitonic solutions with broad edges, Fig.1b
shows an example of the multi-solitonic solutions with sharp edges,
Fig.1c shows an example of the single-solitonic solutions with broad
edges, Fig.1d shows an example of the single-solitonic solutions
with sharp edges, and Fig.1d and 1e show an example of the
multi-solitonic solutions with an envelope.}
\end{table}

For the special choice $\lambda_{1i}=\infty$, the coefficient
$u^+=0$, and the solutions , Eqs.~(\ref{psi_rep}) and
(\ref{gp_attr}), reduce to the simple forms:
\begin{equation}
\psi(x,t)=e^{i\phi_0}\left(A\pm i{\sqrt{8}\lambda_{1r}}\,
e^{-i\beta}{\rm csch}{\theta }\right) \label{psi_rep_simp},
\end{equation}
\begin{equation}
\psi(x,t)=e^{i\phi_0}\left( A-{\sqrt{8}\lambda_{1r}}\,
e^{-i\beta}{\rm sech}{\theta } \right) \label{gp_attr_simp}.
\end{equation}
These two equations show that our solution consists of a bright
soliton embedded in the background \cite{liang}.

\begin{figure}
\begin{center}
\includegraphics[width=10cm]{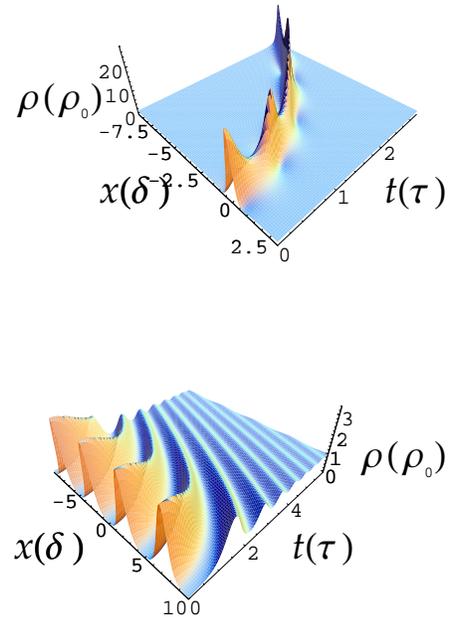}
\end{center}
\caption[a]{Color online Surface plots of the density
$\rho(x,t)=|\psi(x,t)|^2$ versus $x$ and $t$ for the attractive
interactions. The upper plot corresponds to Fig.~1(d) while the
lower figure corresponds to Fig~1(a).} \label{fig2}
\end{figure}

The trajectory of a given soliton peak is obtained from the
condition $\theta=0$ (for single solitons) or  $\beta=0$ (for
multiple solitons). The former gives
$x=-t^2-2(\lambda_{1i}-\lambda_{1r}\Delta_i/\Delta_r)t+\delta_r/\sqrt{2}$
and the latter gives
$x=-t^2-2(\lambda_{1i}+\lambda_{1r}\Delta_r/\Delta_i)t+\delta_i/\sqrt{2}$.
The trajectory is thus parabolic in time with an acceleration of -1.
In real units, this acceleration equals $-F/m$ which is equal to the
acceleration associated with the gravity-like waves propagating on
the surface of the condensate \cite{usama_peth,kett_surf}. This
behavior can be clearly seen in Fig.~\ref{fig2}.

\begin{figure}
\begin{center}
\includegraphics[width=10cm]{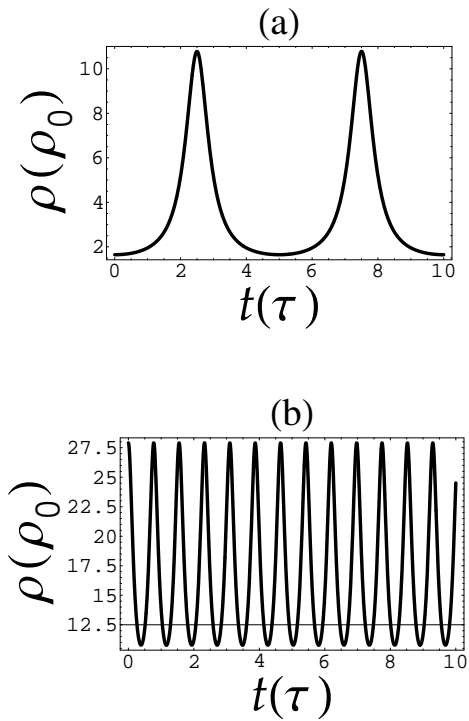}
\end{center}
\caption[a]{The soliton peak density along its trajectory for
attractive interactions. The upper figure corresponds to Fig.~1(c)
and the lower figure corresponds to Fig.~1(d).} \label{fig3}
\end{figure}

Using the above relation between $x$ and $t$, we can eliminate $x$
from $\rho$ to obtain the peak soliton density as a function of
time, i.e., along the trajectory. This is shown in Fig.~\ref{fig3}
where we notice that the soliton peak oscillates between a minimum
and a maximum. The frequency of the oscillation equals
$\sqrt{8}|\Delta|^2\lambda_{1r}/\Delta_r$. For the case of
Fig.~\ref{fig1}(c), where $u^+_r>0$, the maximum appears at times
defined by $\beta=(2n+1)\pi$, namely
$t=n\pi\Delta_r/\sqrt{8}|\Delta|^2\lambda_{1r}$, and the minimum
appears at times defined by $\beta=2n\pi$, $n=0$, 1, 2, $\dots$. The
maximum peak density is given by $(A+\sqrt{8}\lambda_{1r})^2$ and
the minimum peak density is given by $(A-\sqrt{8}\lambda_{1r})^2$.
For the case of Fig.~\ref{fig1}(d), the situation is reversed since
$u^+_r<0$. During this peak oscillation, the number of atoms in the
solitons is being exchanged with the background maintaining a
dynamic stability \cite{liang}. The frequency of atoms exchange is
constant with time, unlike the case of Ref.~\cite{liang}.

\begin{figure}
\begin{center}
\includegraphics[width=10cm]{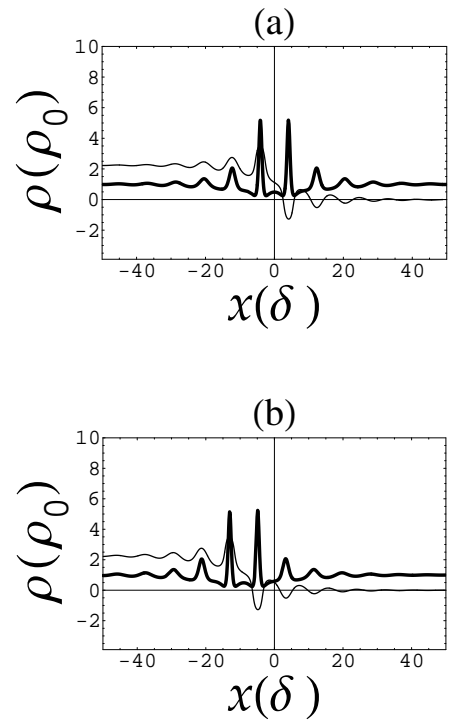}
\end{center}
\caption[a]{The density (solid curve) and phase (light curve) of a
solitonic solution for attractive interactions with
$\delta_r=\delta_i=0$, $A=1$, $\lambda_{1i}=0.03$, and
$\lambda_{1r}=0.6$. The upper plot is for $t=0$ and the lower plot
is for $t=0.85\tau.$} \label{fig4}
\end{figure}

An interesting feature of the soliton is its phase. We plot in
Fig.~\ref{fig4} the phase of a multiple-soliton solution. This shows
that the phase difference between the main neighbouring solitons is
$\pi$. It shows also that this phase difference, the width of
solitons, and the distances between them do not change with time.
The soliton trains realized experimentally in a one-dimensional
condensate with attractive interactions \cite{randy} are very
similar to our solitonic solutions in Fig.~1(d) and (e). It should
be noted here that this is the case since, in addition to describing
the surface of the condensate, Eq.~{\ref{gp2}} describes also a
one-dimensional condensate as long as the density is not too large
\cite{braz}. The center of mass motion of the soliton train in our
case is different than in the experiment of \cite{randy} due to the
fact that we use a linear potential while in the experiment a
quadratic potential is used. On the other hand, the density profile,
relative phase, and number of solitons in the train can, in
principle, be accounted for. This is the case since, in our theory
and in the experiment, the width of the individual solitons (of
order $\delta$) is much less than the that of the background (of
order $R$).

Since the density profile of the exact solitonc solutions found here
depend only on the coordinate perpendicular to the surface of the
condensate, we predict that in a spherical condensate, a
three-dimensional shell-like soliton may exist which in a sense
similar to the ones reported in Refs.~\cite{boris1,boris2}.
Furthermore, the above-described dynamics indicates that, starting
from the surface, this shell will be shrinking in radius.

For the case of repulsive interactions, Eq.~\ref{psi_rep} shows that
the density diverges at certain points along the parabolic
trajectory described above. These points correspond to the points in
Fig~\ref{fig3} where the soliton peak density is maximum. The fact
that the density diverges, does not make this solution nonphysical,
since the number of atoms in the soliton is finite.

It should be mentioned that exact solitonic solutions of
Eq.~(\ref{gp2}) have been essentially obtained using the so-called
inverse-scattering method \cite{liu}. However, the present work represents
another method of obtaining such exact solitonic solutions.
We believe that our method in \cite{usama_darboux} of obtaining the Lax pair
is more systematic since the Lax pair of Ref.~\cite{liu} was introduced as
an assumption.
Furthermore,
in Ref.~\cite{liu}, only formal solutions are derived for the soliton train case.
This is in contrast
with the present work where we obtain explicit single as well as multiple
solitons solutions. Finally, while
in Ref.~\cite{liu}, only the attractive interactions case is considered,
we have derived here solutions for both the attractive and repulsives cases.

In conclusion, we have found exact solitonic solutions of a
time-dependent Gross-Pitaevskii equation with linear trapping
potential for both cases of attractive and repulsive interatomic
interactions. These solutions may be regarded as solitons in the
surface region of a three-dimensional Bose-Einstein condensate, or
solitons in a one-dimensional condensate.



 \end{document}